
\input epsf
\input phyzzx
\hsize=40pc
%
\catcode`\@=11 
\def\NEWrefmark#1{\step@ver{{\;#1}}}
\catcode`\@=12 

\def\square{\kern1pt\vbox{\hrule height 1.2pt\hbox{\vrule width 1.2pt\hskip 3pt
   \vbox{\vskip 6pt}\hskip 3pt\vrule width 0.6pt}\hrule height 0.6pt}\kern1pt}

\def\bra#1{\langle #1 |}
\def\ket#1{| #1 \rangle}

\def\ov{{\overline}}

\def\C{{\cal C}}

\def\K{{\cal K}}

\def\M{{\cal M}}

\def\O{{\cal O}}
\def\P{{\cal P}}

\def\V{{\cal V}}

\def\k{{\kappa}}

\def\ov{\overline}

\def\wh{\widehat}

\def\P{{\cal P}}
\def\V{{\cal V}}
\def\O{{\cal O}}

\singlespace

\def\define#1#2\par{\def#1{\Ref#1{#2}\edef#1{\noexpand\refmark{#1}}}}
\def\con#1#2\noc{\let\?=\Ref\let\<=\refmark\let\Ref=\REFS
         \let\refmark=\undefined#1\let\Ref=\REFSCON#2
         \let\Ref=\?\let\refmark=\<\refsend}

\let\refmark=\NEWrefmark

\define\zwiebachlong{B. Zwiebach, `Closed string field theory: Quantum
action and the Batalin-Vilkovisky master equation', Nucl. Phys {\bf B390}
(1993) 33, hep-th/9206084.}

\define\bergmanzwiebach{O. Bergman and B. Zwiebach, `The dilaton theorem
and closed string backgrounds', to appear in Nucl. Phys. B, hep-th/9411047.}

\define\polchinski{J. Polchinski, `Factorization of bosonic string amplitudes',
Nucl. Phys. {\bf B307} (1988) 61.}

\define\belopolskyzwiebach{A. Belopolsky and B. Zwiebach,
`Off-shell string amplitudes: Towards a computation
of the tachyon potential',  to
appear in Nucl. Phys. B, hep-th/9409015.}

\define\alvarez{L. Alvarez-Gaume, C. Gomez, G. Moore and C. Vafa,
`Strings in the operator formalism', Nucl.
Phys. {\bf B303} (1988) 455;\hfill\break
C. Vafa, `Operator formulation on Riemann surfaces',
Phys. Lett. {\bf B190} (1987) 47.}

\define\senzwiebachtwo{A. Sen and B. Zwiebach, `Quantum background
independence of closed string field theory', Nucl. Phys. {\bf B423} (1994) 580,
hep-th/9311009.}

\define\whittakerwatson{E. T. Whittaker and G. N. Watson,
{\it A course in Modern
Analysis}, Cambridge University Press, 1973.}

\define\leclair{A. LeClair, M. E. Peskin and C. R. Preitschopf, Nucl. Phys.
{\bf B317} (1989) 411.}

\singlespace
{}~ \hfill \vbox{\hbox{MIT-CTP-2448}
\hbox{hep-th/9507038}\hbox{
} }\break
\title{VACUUM VERTICES AND THE GHOST-DILATON}
\author{Sabbir Rahman  \foot{E-mail address: rahman@mitlns.mit.edu }
and Barton Zwiebach \foot{E-mail address: zwiebach@irene.mit.edu
\hfill\break Supported in part by D.O.E.
cooperative agreement DE-FC02-94ER40818.}}
\address{Center for Theoretical Physics,\break
Laboratory of Nuclear Science\break
and Department of Physics\break
Massachusetts Institute of Technology\break
Cambridge, Massachusetts 02139, U.S.A.}

\abstract
{We complete the proof of the ghost-dilaton theorem in string theory by
showing that the coupling constant dependence of the vacuum vertices
appearing in the closed string action is given correctly by one-point
functions of the ghost-dilaton. To prove this at genus one we develop the
formalism required to evaluate off-shell amplitudes on tori.}
\endpage

\singlespace

\chapter{Introduction and Summary}

Whenever a string background contains a nontrivial ghost-dilaton state,
a shift of the string field along this state is expected to
alter the value of the dimensionless string coupling constant.
This expectation, the so-called ``ghost-dilaton theorem'',
was recently established in Ref.[\bergmanzwiebach]
as a property of the  covariant closed string field theory action.
More precisely, this work established the result only for the field-dependent
terms in the string action. In this brief paper we complete the proof of the
ghost-dilaton theorem by showing that a shift by the ghost-dilaton also
changes the value of the string coupling  in the field-independent terms of
the string action. These field-independent terms arise as vacuum
vertices of nonvanishing
genus $g$.

The effect of the ghost-dilaton shift is found by inserting a ghost-dilaton
on the vertices defining the string action. For each vertex,
one integrates over the corresponding moduli space of surfaces
the result of integrating a ghost-dilaton insertion over each surface.
For string vertices {\it with}
external punctures it was found convenient to place
the antighost insertions for moduli changes
on an external puncture. This choice makes it clear
that a suitable ghost-dilaton
insertion computes the Euler number of the bordered
surfaces comprising the string
vertex. Complications arise for the
vacuum vertices, though the nature of these difficulties are different
for the cases $g\geq 2$ and the case $g=1$. In general,
there are no external punctures available to support the moduli-changing
insertions. Moreover, for genus
one, the existence of conformal isometries dictates that there is no integral
over the position of the dilaton insertion.

In section \S2 we demonstrate how the methods of Ref.[\bergmanzwiebach]
actually apply for $g\geq 2$ even though the
moduli-changing Schiffer vectors are supported on the puncture where the
dilaton is inserted. We find that integrating the dilaton over
a vacuum vertex
correctly reproduces the Euler characteristic
of the unpunctured surfaces.

In section \S3 we discuss the
evaluation of general off-shell amplitudes for punctured tori. This
is necessary for the ghost-dilaton, since despite being a physical
state it has no primary BRST representative.
We construct any once-punctured torus with an arbitrary local
coordinate around the puncture  by sewing two punctures of
a particularly simple three-punctured sphere
with an appropriately chosen
sewing parameter. We  give explicit expressions for the
Schiffer  vectors which generate the tangents to
the moduli space of punctured tori with local coordinates. The
results of this section represent an extension to genus one of
the techniques developed in Ref.[\belopolskyzwiebach].

In section \S4 the above ideas are used
to show that both the canonical string two-form in the direction of the
ghost-dilaton $\ket{D}$ {\it and} the canonical string one-form
in the direction
of $\ket{\chi} = -c_0^-\ket{0}$ vanish identically on the moduli space
of once-punctured tori (here $\ket{D} = -Q \ket{\chi}$).
These results imply that the genus one
field-independent terms in the string action are
unchanged by a shift of the ghost-dilaton.

\chapter{Vacuum Vertices of Genus $g\geq 2$}

In the following we assume some familiarity with the notation of
[\bergmanzwiebach], which should be consulted for definitions.
If a shift of the ghost-dilaton is to change the coupling constant
in the $g\geq 2$ field-independent terms of the string action the
following equation should hold (eqn.~(8.14) of [\bergmanzwiebach])
$$f_D(\ov\K\V_{g,0})=(2g-2)f(\V_{g,0})\,,\qquad g\geq 2\,.\eqn\raju$$
The left hand side denotes a ghost-dilaton insertion over the vacuum vertex
of genus $g$, and the right hand side is (minus)
the Euler number of a genus $g$ surface times the vacuum vertex.
The purpose of the present section is to provide a proof of this equation.

The right hand side involves the integration over
$\V_{g,0}\subset \M_{g,0}$, of a $(6g-6)$-form
defined in eqn.~(3.15) of ref.[\senzwiebachtwo]
$$f(\V_{g,0}) \equiv {\cal N}^{3-3g}\int_{\V_{g,0}}d\,{\vec \xi}\,\,
\bra{\wh\Sigma\,}{ b}(\wh{ v}_{\xi_1})\cdots{ b}(
\wh{ v}_{\xi_{6g-6}})\ket{0}\,.\eqn\karabi$$
In this equation ${\cal N} \equiv -2\pi i$, and
$d{\vec \xi} \equiv d\xi_1\wedge\cdots\wedge\xi_{6g-6}$, where
$(\xi_1,\cdots ,\xi_{6g-6})$ is a set of coordinates in $\V_{g,0}$.
Given a surface $\Sigma \in \V_{g,0}$, $\wh\Sigma\in\wh\P_{g,1}$ denotes
the same surface, with the addition of an extra puncture equipped with
a local coordinate (defined up to a phase).
In addition, $\wh{v}_{\xi_i}$ is the Schiffer vector associated to the
tangent $\partial /\partial \xi_i~$. More precisely, it represents
a tangent in $T_{\wh\Sigma}\wh\P_{g,1}$ chosen to project down to
$\partial/\partial\xi_i\in T_{\Sigma}\M_{g,0}$
upon deletion of the extra puncture.
While the ingredients used to
build the form integrated
in \karabi\ depend on position and coordinates around the puncture, as well
as a choice of representatives for the tangents, the resulting
form is independent of this data [\alvarez].

Let us elaborate on the  ambiguity in the choice of  Schiffer vectors
representing the tangents in \karabi.
If the local coordinate at the puncture
is denoted as $z$,
Schiffer vectors $v'(z) =a_0 + a_1 z + \cdots$, regular
at the puncture, can only change the position
and local coordinates at the puncture. Since  the above
form is independent of this data, the change
${b}(\wh{ v}_{\xi_i})\to { b}(\wh{ v}_{\xi_i})+
{ b}( v')$ should make no difference in \karabi. This follows immediately
from ${b}( v')\ket{0}=0$, which, in turn, holds since
$b_{n\geq -1}$ and $\bar b_{n\geq -1}$
annihilate the vacuum.

Consider now the left hand side of \raju . Here we must integrate over
the position of an extra puncture on
each surface in the space $\V_{g,0}$. Since the ghost-dilaton is not primary
we need a a family of local coordinates throughout each surface of $\V_{g,0}$.
Such a family is obtained by introducing, continuously over $\V_{g,0}$,
a conformal metric on every surface. This metric is used to
define a family of local coordinates
via the prescription
of Ref.[\polchinski], as elaborated in  \S 3.1 of [\bergmanzwiebach]. We now
write the left hand side of \raju\ as follows
$$f_D(\ov\K\V_{g,0})
\equiv {\cal N}^{2-3g}\int_{\ov\K\V_{g,0}}d\,{\vec\xi}
\wedge d\lambda_1\wedge d\lambda_2
\cdot\bra{\wh\Sigma\,} b({ v}_{\xi_1})\cdots b
({v}_{\xi_{6g-6}})\,{b}({ v}_{\lambda_1}){ b}({ v}_
{\lambda_2})\ket{D}\,.\eqn\shuja$$
The $\xi$ are coordinates $\M_{g,0}$, the same coordinates used
 in \karabi.
The real parameters $(\lambda_1,
\lambda_2)$ parameterize the position of the dilaton puncture. The Schiffer
vectors $v_{\lambda_i}$ are chosen to alter only the
position and coordinate data for the dilaton puncture, while the
$v_{\xi_i}$ represent the tangents that change the moduli of the surface,
and possibly the data at the puncture. The vectors $v_{\xi_i}$ differ from
the vectors $\wh v_{\xi_i}$ appearing in \karabi\ by vectors $v'_i$ regular
at the puncture. They also may differ by irrelevant ``Borel vectors"
$\tilde v$. Such vectors satisfy $\bra{\wh\Sigma} b(\tilde v) = 0$.

We now recall from Ref.[\bergmanzwiebach] \S5.2 that:
$d\lambda_1\wedge d\lambda_2\,{b}({ v}_{\lambda_1}){b}({ v}_
{\lambda_2})\ket{D} = i R^{(2)} (\rho) \ket{0}$, where
$R^{(2)} (\rho)$ is the curvature two-form associated to the conformal
metric $\rho$ used to extract the coordinates for the ghost-dilaton
insertion.
Replacing this into \shuja\ we find
$$f_D(\ov\K\V_{g,0})
\equiv -{\cal N}^{3-3g}\int_{\V_{g,0}}d\,{\vec\xi}\int_\Sigma
\cdot\bra{\wh\Sigma\,}{b}(\wh v_{\xi_1})\cdots{b}
(\wh v_{\xi_{6g-6}})\ket{0}\, {1\over 2\pi} R^{(2)}(\rho) \,,\eqn\shujaa$$
where we used the remarks below eqn.\karabi\ to replace the vectors
$v_{\xi_i}$ by the vectors $\wh v_{\xi_i}$. As emphasized earlier, the
form $d{\vec \xi}\,\bra{\wh\Sigma\,}{ b}(\wh v_{\xi_1})\cdots{b}
(\wh v_{\xi_{6g-6}})\ket{0}$ is independent of the position and coordinates
of the puncture placed on $\Sigma$. Therefore, the integral
over $\Sigma$ in \shujaa\
can be readily evaluated to give ${1\over 2\pi} \int_\Sigma R^{(2)}=2-2g$.
Back in \shujaa\ we
find
$$f_D(\ov\K\V_{g,0})
= (2g-2)\, {\cal N}^{3-3g}\int_{\V_{g,0}}d\,{\vec\xi}
\bra{\wh\Sigma\,}{b}(\wh v_{\xi_1})\cdots{ b}
(\wh v_{\xi_{6g-6}})\ket{0}\,  \,.\eqn\donne$$
Comparison with \karabi\ immediately gives the desired result.

\chapter{Off-Shell Amplitudes in Tori}

In this section, we use a representation of the punctured torus in terms of a
sewn three-punctured sphere to find the form of the Schiffer vectors which
independently generate modulus deformations and local coordinate changes.
We then use these to obtain explicit expressions for the canonical forms
over the moduli space of tori and in doing so set up for the formalism
required to evaluate off-shell amplitudes in tori.

\section{Once-punctured tori from three punctured spheres}

Let $\bra{\Sigma_1 ; w_1}$ and $\bra{\Sigma_2 ; w_2}$ denote the surface
states corresponding to the punctured Riemann surfaces $\Sigma_1$ and
$\Sigma_2$. We  single out a puncture on each surface, labelled
puncture one and puncture two, and having local coordinates $w_1$ and $w_2$,
respectively. Let $\Sigma (q)$ denote the  surface produced by
sewing together these surfaces with sewing parameter $q$. The surface
state corresponding to the sewn surface is given by,
$$\bra{\Sigma (q)} =   \bra{\Sigma_1; w_1\,} \bra{\Sigma_2; w_2}
q^{L_0^{(1)}} \bar q^{\overline L_0^{(1)}} \ket{R_{12}} \,, \quad
\Sigma (q) \equiv\Sigma_1\cup_q\, \Sigma_2 : w_1w_2=q\, \,.\eqn\eleven$$
Note that the second state space could equally well be used
for the Virasoro operators. This follows from the exchange
symmetry of $\ket{R_{12}}$.

Consider now the canonical two punctured sphere $z_1 (z) = z$, $z_2 (z) = 1/z$,
with an additional puncture at $z=1$ whose local coordinate $w$
will be defined as,
$$ w = {1\over 2\pi i} \ln z \,  .\eqn\onep$$
This three punctured sphere will be denoted by $R_{123}$, and the
corresponding surface state  by $\bra{R_{123}}$.
The sphere is described in the $w$-plane as the cylindrical
region determined by the identification:
$w \sim w + 1$.

Suppose we now sew the coordinates $z_1$ and $z_2$ via the
identification
$z_1 z_2 = q$, where $q \equiv \exp (2\pi i \tau )$.
In terms of the $z$ uniformizer this means identifying according to
$z\sim  z \, \exp (2\pi i \tau )$, which in $w$ coordinates reads
$w \sim w + \tau\, .$
It follows from the identifications
that we have obtained a torus with Teichm\"uller parameter
$\tau$. The surface state $\bra{\Sigma (\tau) ; w}$
describing this once-punctured
torus may therefore be written as,
$$\bra{\Sigma (\tau) ; w} = \bra{R_{123}}
q^{L_0^{(1)}} \bar q^{\overline L_0^{(1)}} \ket{R_{12}}\,.\eqn\sixp$$
This is the once-punctured torus $w\sim w+1$, $w\sim w+\tau$,
with local coordinate $w$ at the puncture $w=0$.

We now consider describing a general family of once-punctured
tori. The local coordinate
at the puncture will depend on the modulus of the torus.
Every torus  will be explicitly realized
in the $w$-plane via the identifications
$w\sim w + 1$ and $w\sim w + \tau$, and the puncture will be
at $w=0$. As we change the modulus, the local coordinate at this
puncture is conveniently described  by a function showing how the unit
coordinate disk $|\xi| \leq 1$
embeds into the torus,
$$w = h_{\tau, \bar\tau} ( \xi) = a(\tau , \bar \tau)\, \xi +
\half b (\tau , \bar \tau )\, \xi^2 + \cdots \eqn\oneppp$$
Note that the function need not depend holomorphically on
the modulus $\tau$. The above equation may be inverted to give the local
coordinate as a function of the uniformizer $w$ of the torus,
$\xi = \xi (\tau,\bar \tau , w)$. For brevity, we will write
$\xi = \xi_\tau ( w)$, and the $w$ dependence will be left implicit
in some cases.

Following \sixp, the
surface state for a torus of modulus $\tau$ having coordinate
$\xi_\tau$ at the puncture is given by,
$$ \bra{\Sigma(\tau) ; \xi_\tau } = \bra{R_{123}^{\xi_\tau}\, }
 q^{L_0^{(1)}} \bar q^{\overline L_0^{(1)}}
\ket{R_{12}} \,,  \eqn\twoppp$$
where the three punctured sphere $R_{123}^{\xi_\tau}$
is the canonical two-punctured sphere $R_{12}$ equipped
with an extra puncture at $z=1$:
$$R_{123}^{\xi_\tau}:\, \quad
 z_1(z) =z , \quad z_2 (z) = 1/z\,,\quad z_3(z) = \xi_\tau (w(z)\,)\,.
\eqn\threeppp$$
Here $w$ is the coordinate
defined in \onep.
The choice of coordinate around the third puncture of the sphere ensures
that the torus will have the desired local coordinate $\xi_\tau(w)$ at the
puncture.

\section{Forms on the moduli space of punctured tori}

Our aim here is to write
the canonical string forms on the moduli space $\P_{1,1}$ of
once-punctured tori. The general case
becomes clear once we  write forms on the
subspace of $\P_{1,1}$ parametrized
by \oneppp\ which defines the local coordinate at the puncture
as a function of the modulus of the torus. This is the case because
the most general tangent to $\P_{1,1}$ is a tangent that changes the
modulus of the torus and adjusts the local coordinate at the puncture.

We consider a torus of modulus $\tau$
and write the one-form and two-forms as
$$\eqalign{
\bra{\Omega^{[1]1,1}(\tau)} &= {\cal N}^{-1}\,
\bra{R_{123}^{\xi_\tau}}
q^{L_0^{(1)}} \bar q^{\overline L_0^{(1)}} \ket{R_{12}} \,
\Bigl[ d\tau  b^{(3)}\Bigl({\partial\over \partial \tau}\Bigr)
+d\bar\tau\, b^{(3)}
\Bigl({\partial\over \partial \bar\tau}\Bigr)\Bigr]\,,\cr
 \bra{\Omega^{[2]1,1}(\tau)} &= {\cal N}^{-1}\, \bra{R_{123}^{\xi_\tau}}
q^{L_0^{(1)}} \bar q^{\overline L_0^{(1)}} \ket{R_{12}} \,\,
d\tau\wedge d\bar\tau\,\, b^{(3)}\Bigl({\partial\over \partial \tau}\Bigr)
\,b^{(3)}\Bigl({\partial\over \partial \bar\tau}\Bigr)\, ,\cr}\eqn\onepp$$
where ${\cal N}= -2\pi i$ is a normalization factor, and the surface
state representing the once punctured torus was built by sewing.
As usual, the antighost insertions
are located at the free puncture. The vectors $\partial/\partial\tau$
and $\partial/\partial\bar\tau$ denote the Schiffer vectors
representing the deformation of the modulus together with the corresponding
deformation of the local coordinates. Since surface states for punctured
spheres are well-known [\leclair, \belopolskyzwiebach],
the  construction of the above forms only
requires the explicit expressions for the Schiffer vectors for the family
of tori described in \oneppp, and the prescription to build the corresponding
antighost insertions.

Let us first consider the case where regardless
of the modulus of the torus, the
local coordinate at $w=0$ is always taken to be the coordinate
$w$. In this case we say, with a slight abuse of language,
that the coordinate does not change as we change
the modulus.
The Schiffer vectors must not  deform
the local coordinate at the free
puncture;~in terms of the three-punctured sphere, the torus deformation
only entails a change of sewing parameter for the sewing of the
first two punctures.
It is therefore convenient to place
the antighost insertions on an interior puncture, say puncture one.
The two-form then reads
$$ \bra{\Omega^{[2]1,1}(\tau)} = {\cal N}^{-1}\, \bra{R_{123}}
q^{L_0^{(1)}} \bar q^{\overline L_0^{(1)}}\, b^{(1)}\Bigl({\partial\over
\partial \tau}\Bigr)
b^{(1)}\Bigl({\partial\over \partial \bar\tau}\,\Bigr)
\ket{R_{12}} \,
d\tau\wedge d\bar\tau\,\, .\eqn\twopp$$
In this presentation, it is a standard calculation to show that,
$$ b \Bigl( {\partial \over \partial \tau} \Bigr)
= -2\pi i b_0 \,, \quad\quad
    b \Bigl( {\partial \over \partial \bar\tau} \Bigr) =
2\pi i \bar b_0\,,\eqn\auxx$$
which can now be used to write
$$ \bra{\Omega^{[2]1,1}(\tau)} = -{\cal N}
\cdot \bra{R_{123}}
q^{L_0^{(1)}} \bar q^{\overline L_0^{(1)}}\, b_0^{(1)}
\bar b_0^{(1)}
\ket{R_{12}} \,
d\tau\wedge d\bar\tau\,\, .\eqn\eightpp$$

\noindent
\underbar{The Schiffer vectors on a torus}.
We can now discuss Schiffer vectors $v(w)$ defined on the neighborhood
of the puncture in the once-punctured torus.\foot{For a pedagogic discussion
of general properties of Schiffer vectors and Schiffer
variations see Ref.[\zwiebachlong].}
In the brief analysis which follows, we shall consider the possible vectors
according to their behavior near the puncture at $w=0$. Let us begin by
considering those vectors fields which vanish at the puncture,
$$v(w) = w^n +  \O (w^{n+1})\,,\quad  n\geq 1 \, . \eqn\oneschiff$$
Such objects do not extend holomorphically throughout the torus since
no bounded elliptic function exists.
Since they extend all the way inwards to $w=0$, where they
vanish, their effect is to change the local coordinate at the puncture.
Corresponding to \oneppp, the explicit form of the Schiffer vectors that
change the local coordinates as $\tau$ changes, can be read from eqn.(6.10) of
Ref.[\belopolskyzwiebach]
$$v_{\tau} (w) = - {\partial h\over \partial \tau}
(\xi (w)) \,, \quad v_{\bar\tau} (w) = - {\partial h\over \partial \bar\tau}
(\xi (w))\,.\eqn\bins$$
These vectors are clearly of the type indicated in \oneschiff.

The constant
vector $v(w) = 1$ is well defined throughout the torus
and therefore changes neither the modulus nor the local coordinate at
the puncture. The  case of a vector with first order pole is more
interesting:
$$v(w) = {1\over w} + \O (w^0) \,\,. \eqn\fivesch$$
This vector cannot be extended analytically throughout the torus as no
elliptic function with a single first order pole exists. Such a vector field,
which neither extends inwards nor outwards, must change the modulus. We are
especially interested in finding the vector field which changes the modulus
without a corresponding change in local coordinate at the puncture.
To this
end, it is useful to consider the logarithmic derivative of the Jacobi theta
function $\theta_1(w|\tau)$,
$$ u (w\,| \tau)\equiv {\theta'_1 ( w \,| \tau)   \over \theta_1 ( w\, | \tau
)}
  ={1\over w} +
\cdots\,,\eqn\sixsch$$
where the prime denotes differentiation with respect to the first
argument. (We follow the conventions of Ref.[\whittakerwatson]; note however
that there $q= e^{i\pi \tau}$). This logarithmic derivative
has  simple  properties under translations,
$$u ( w + \pi \,| \tau) = u ( w \,| \tau)\,, \quad\quad
u ( w + \pi\tau \,| \tau) = -2i + u ( w \,| \tau) \,.\eqn\thetaf$$
For our purposes, it is convenient to choose the particular Schiffer vector,
$$v_o(w) ={i\over 2}\, u (\pi w \,| \tau ) \,,\eqn\desvec$$
which, by virtue of \thetaf\ satisfies,
$$v_o(w+1) = v_o(w)\,,\quad v_o(w+\tau) =1 + v_o(w) \,.\eqn\yesp$$
As will be seen later, this choice of Schiffer vector is tailored precisely to
change the modulus without changing the local coordinate at the puncture
(again, in the sense that its dependence on the uniformizer is unaltered).
Since
$\theta$-functions are entire, the only pole of $v_o$  is at $w=0$.

To conclude our analysis, consider
vectors of the form,
$$v(w) = {1\over w^n} + \O (w^{1-n}) \,,\quad n\geq 2\,. \eqn\higher$$
Vectors with such leading behavior can always be reduced to one of the cases
already discussed by subtracting a suitable linear combination of elliptic
funtions. In particular the Weierstrass $\P$ function may be used to
eliminate a second order pole, while its derivatives can be used to
eliminate higher order poles.

\noindent
\underbar{The Antighost Insertions}.
We now describe the antighost insertions corresponding to the
Schiffer vectors discussed above. We treat separately the
contributions from vectors that change coordinates only and from
vectors that only change moduli. Finally, the
results are combined to give the expression of a general form on the
moduli space of once-punctured tori.

The antighost insertions corresponding to the coordinate changing
Schiffer vectors in \bins\ are given by
$$\eqalign{
b (v_\tau) &= \oint {dw\over 2\pi i} b(w) v_\tau (w) +
\oint {d\bar w\over 2\pi i}\, \bar b (\bar w)\,
\overline{v_{\bar\tau}(w)}\,,\cr
b (v_{\bar\tau}) &= \oint {dw\over 2\pi i} b(w) v_{\bar\tau} (w) +
\oint {d\bar w\over 2\pi i} \bar b (\bar w) \overline{v_\tau (w)}\,,\cr}
\eqn\oog$$
where overbar on a vector denotes complex conjugation, and we integrate using
$\oint dw/2\pi iw = \oint d\bar w/ 2\pi i\bar w =1$.
In these equations $b (v_\tau)$ and $b (v_{\bar\tau})$ are simply the
portions of $b ( \partial/\partial \tau)$ and of
$b ( \partial/\partial \bar\tau)$ representing the coordinate changes
due to the change in modulus. The explicit oscillator
form of the antighost insertions follows readily from the above equations
once the Schiffer vectors in \bins\ are written as power series in $w$.
Note that the insertions are acting on the external puncture, but use the
$w$ uniformizer. This is convenient when it comes
to re-expressing them in terms of the first and second state spaces,
the spaces that are traced over.
Since $b$ is primary, \oog\ is coordinate independent and the
insertions can be expressed in terms of oscillators $b^{(\xi)}$ by using
the Schiffer vectors referred to the $\xi$ coordinates. This may be useful
since the external state is written in terms of such oscillators.

Let us now consider the antighost insertion corresponding to a change in
modulus only. The vector field we need is precisely
the vector $v_o(w)$
introduced in \desvec. If we define
$$b(v_o) = \oint {dw\over 2\pi i} b(w) v_o(w)\,,\quad
b( \overline {v_o}) = \oint {d\bar w\over 2\pi i}\bar b( \bar w)
\overline{ v_o(w)},\eqn\nineev$$
then, the correctness of our claim requires that
$$\bra{R_{123}^{\xi_\tau}} \,
q^{L_0^{(1)}} \bar q^{\overline L_0^{(1)}} \ket{R_{12}}\,
 b(v_o) =
\bra{R_{123}^{\xi_\tau}} \,(-2\pi i\,b_0^{(1)})\,
q^{L_0^{(1)}} \bar q^{\overline L_0^{(1)}} \ket{R_{12}}\,. \eqn\sevenev$$
This equation relates an antighost insertion on the puncture to an
antighost operator inside the trace and by virtue of
\auxx\  it justifies the assertion that
the antighost insertion $b(v_o)$ describes a change of modulus only, without
affecting the local coordinate at the puncture. The proof of this identity
is given in the appendix. In a similar
manner,
$$\bra{R_{123}^{\xi_\tau}} \,
q^{L_0^{(1)}} \bar q^{\overline L_0^{(1)}} \ket{R_{12}}\,
b(\overline {v_o}) =
\bra{R_{123}^{\xi_\tau}} \,(2\pi i\,\bar b_0^{(1)})\,
q^{L_0^{(1)}} \bar q^{\overline L_0^{(1)}} \ket{R_{12}}\,. \eqn\eightev$$

Our results are now complete. The antighost insertions
representing {\it both} changes of moduli and local coordinates are
given as
$$b\,\Bigl({\partial\over \partial \tau}\Bigr) =  b(v_o) + b (v_\tau)\,,
\quad \quad
b\,\Bigl({\partial\over \partial \bar\tau}\Bigr) =
 b(\overline{v_o}) + b(v_{\bar\tau})\,. \eqn\contrib$$
These expressions can now be substituted back in \onepp\ and give
the explicit expressions for the canonical string forms on the moduli
space of once-punctured tori.

\chapter{The Case of Genus One}

In this section, we apply the earlier results to prove the ghost-dilaton
theorem at genus one. According to
[\bergmanzwiebach], the theorem is proven if the following equation holds,
$$f_D(\underline\V_{1,1})-f_\chi(\Delta\underline\V_{0,3})=\k{d\over d\k}
\left(S_{1,0}+{1\over 2}\ln\rho\right)\,.\eqn\ridoy$$
The expression in parenthesis in the right hand side corresponds
to the elementary contribution to the one-loop free energy appearing in
the string action. It
is expected to be independent of the string
coupling $\kappa$, so ideally one would hope that the left hand side
of the equation vanishes. It does. In what follows, we verify that in fact both
terms appearing in this left hand side vanish independently.

\noindent
\underbar{Preliminary remark}. We shall consider first $f_\chi(\Delta
\underline\V_{0,3})$, with $\ket{\chi} = -c_0^-\ket{0}$,
$$f_\chi(\Delta\V_{0,3}) =  \int_{\Delta\V_{0,3}}{d\theta\over 2\pi}\,
\bra{V_{123}}
b_0^{-(1)}e^{i\theta L_0^{-(1)}}\ket{R_{12}}c_0^{-(3)}\ket{0}_3\,.\eqn\yamin$$
The geometrical interpretation is that we take our choice of three-string
vertex (denoted here by $\bra{V_{123}}$), and twist-sew two of the
punctures, inserting the state $\ket{\chi}$ at the third puncture. It happens
that certain simplifications occur in the special case where the three-string
vertex is chosen to be the Witten vertex $\bra{W_{123}}$.
This example gives
an insight into how the expression may be evaluated for the general case.

The $\bra{W_{123}}$ vertex satisfies the conservation law:
$\bra{W_{123}}(c_0^{(1)}+ c_0^{(2)}+c_0^{(3)} )=0$,
which allows us to replace the $c_0^{-(3)}$ in \yamin\
by the factor $(c_0^{-(1)}+c_0^{-(2)})$ immediately to the right of the
surface state. This factor annihilates the reflector $\ket{R_{12}}$,
and all that remains is the  term picked up by anticommutation
with $b_0^{-(1)}$,
$$f_\chi(\Delta\V_{0,3})= \int_{\Delta\V_{0,3}}
{d\theta\over 2\pi}\, \bra{W_{123}}e^{i\theta
L_0^{-(1)}}\ket{R_{12}}\,\ket{0}_3\, .\eqn\evana$$
Let us consider the integrand of this expression. We
have a three-punctured sphere with two punctures
sewn together, and a state (in this case the vacuum) inserted in the last
puncture.
Such an object is a once-punctured torus of some modulus
$\tau(\theta)$. The calculation of $\tau$ as a function of $\theta$ is
in general nontrivial, but will not be necessary for our arguments.
Referring back to our
discussion of \S3.1, any once-punctured torus with arbitrary modulus
$\tau$ can be made by sewing
together the punctures of $R_{123}^{\xi_\tau}$ with an
appropriately chosen sewing parameter, and local coordinate $\xi_\tau$.
We may therefore
rewrite \evana\ as,
$$f_\chi(\Delta\V_{0,3})= \int_{\Delta\V_{0,3}}
{d\theta\over 2\pi} \bra{R_{123}^{\xi_\tau}}
q^{L_0^{(1)}}
\ov q^{\ov L_0^{(1)}}\ket{R_{12}}\,\ket{0}_3\,=
  \int_{\Delta\V_{0,3}}{d\theta\over 2\pi} \bra{R_{12}}
q^{L_0^{(1)}}\ov q^{\ov L_0^{(1)}}\ket{R_{12}}\,,\eqn\laboni$$
where $q\equiv e^{2\pi i\tau(\theta)}$.
It becomes clear that we are actually dealing with the supertrace of
the operator $q^{L_0}\ov q^{\ov L_0}$\foot{One can readily show that
$\bra{R_{12}}A^{(1)} \ket{R_{12}} = \sum_s (-)^s\bra{\Phi^s} A \ket{\Phi_s}
\equiv \hbox{str}( A)$}.

The supertrace of any operator
vanishes unless the operator explicitly contains the factor
$b_0c_0\bar b_0\bar c_0$ . The necessity of the factor $b_0c_0$
follows by considering the holomorphic sector.
The state space splits into two  sectors, identical except for the fact
that they are  built upon the vacua $\ket{0}$ and $c_0\ket{0}$ respectively.
The different fermion numbers of these two sectors means that their
contributions to the supertrace cancel. It is therefore necessary
to project out one of these sectors while still conserving ghost number,
the only operator suitable for this being $b_0c_0$. Similarly,
the antiholomorphic sector requires the factor  $\bar b_0\bar c_0$.

Since neither $L_0$ nor $\ov L_0$  contain ghost zero modes, the integrand in
\laboni\ vanishes identically. Note that such simple arguments could not
be made for \evana. Upon deletion of the third puncture, the leftover
two-punctured sphere does not coincide with $R_{12}$ and the absence of
ghost zero modes is not manifest.

\noindent
\underbar{The general case.}
We now consider to the case of the
general three-string vertex. Since $f_\chi (\Delta \V_{0,3})$
is simply an integral of the canonical string one-form
along a one-parameter subspace of $\P_{1,1}$, the  considerations
of the previous section indicate that it can be written as
$$f_\chi(\Delta\V_{0,3})={\cal N}^{-1}
\hskip-6pt\int_{\Delta \V_{0,3}}\bra{R_{123}^{\xi_\tau}}
q^{L_0^{(1)}}\ov q^{\ov L_0^{(1)}}\ket{R_{12}}\Bigl[
d\tau b^{(3)}\Bigl({\partial\over
\partial \tau}\Bigr)
+d\ov\tau\, b^{(3)}\Bigl({\partial\over
\partial \bar\tau}\Bigr)\Bigr] \ket{\chi}_3\,,\eqn\rinku$$
Here the tori are built by gluing two punctures of the $\tau$-dependent
$R_{123}^{\xi_\tau}$ spheres, where the coordinates at the third puncture
are determined by the desired coordinates on the punctured tori
$\Delta\V_{0,3}$.
The antighost insertions represent both changes of moduli, and changes
in the local coordinates as
outlined in detail in the previous section (see \contrib).

An important property of the $R_{123}^{\xi_\tau}$ sphere is that
the local coordinates at punctures one and two reach puncture three.
This means that
we can, by way of conformal transformations, map any operator acting on
the third state space to an operator acting on the first or on the
second state space.
Such  mapping couples neither left-movers with right-movers nor ghosts with
antighosts. Exploiting this fact
to our advantage, we use conformal mapping to transfer all oscillators
in the third puncture to the
first puncture. Once again, the vacuum state deletes the special puncture of
$\bra{R_{123}^{\xi_\tau}}$, and what remains is an expression of the form,
$$f_\chi(\Delta\V_{0,3})\sim \int_{\Delta \V_{0,3}}\bra{R_{12}}\left(d\tau
(b^{(1)}+\ov b^{(1)}) +d\ov\tau(b^{(1)}+\ov b^{(1)})\right)(c^{(1)}-\ov
c^{(1)})q^{L_0^{(1)}}\ov q^{\ov L_0^{(1)}}\ket{R_{12}}\,,\eqn\rimi$$
where each ghost or antighost term  indicated schematically above,
is in general a sum of many modes.
For our present purposes we  only need to know
the holomorphic or antiholomorphic
characted of a given term. This is displayed by the absence or presence
of an overbar.
It is now quite clear that at most two of
the required four ghost zero modes may be present in any one term. The
integrand
therefore vanishes identically.

We can now apply the same ideas to
$f_D(\underline\V_{1,1})$
$$\eqalign{f_D(\underline\V_{1,1})
&= {\cal N}^{-1}\hskip-6pt\int_{\underline\V_{1,1}}d\tau\wedge d\ov\tau\,
\bra{R_{123}^{\xi_\tau}}q^{L_0^{(1)}}\ov
q^{\ov L_0^{(1)}}\ket{R_{12}}
\,b^{(3)}\Bigl({\partial\over\partial \tau}\Bigr)
\, b^{(3)}\Bigl({\partial\over\partial \bar\tau}\Bigr)\ket{D}_3 \cr
&\sim\int_{\underline\V_{1,1}}d\tau\wedge d\ov\tau\,\bra{R_{12}}(b^{(1)}+
\ov b^{(1)})(b^{(1)}+\ov b^{(1)})(c^{(1)}c^{(1)}-\ov
c^{(1)}\ov c^{(1)})q^{L_0^{(1)}}\ov q^{\ov L_0^{(1)}}\ket{R_{12}}\,,}\eqn
\jhumka$$
where once again we used conformal mapping to move all oscillators to the
first puncture (recall $\ket{D} = (c_1c_{-1}-\bar c_1\bar c_{-1})\ket{0} $).
This time the expression vanishes due
to the absence of the combination $c_0 \ov c_0$.
We have thereby shown that the left hand side of \ridoy\ vanishes. This
concludes our proof of the ghost-dilaton theorem for the field independent
terms in the string action.

\ack
We would like to thank A. Belopolsky and R. Dickinson for their helpful
comments
during the course of this work.

\appendix

\noindent
We consider here the proof of \sevenev.
The starting point is
$$\bra{R_{123}^{\xi_\tau}} \oint_{\C_0} {dw\over 2\pi i} b(w) v_o(w) \,
q^{L_0} \bar q^{\overline L_0} \ket{R_{12}}\,, \eqn\beginev$$
where the integral is around the point $w=0$.
Note that the antighost insertion is
written in terms of the $w$-coordinate at the puncture.
This expression is viewed as an integral on the sphere $R_{123}^{\xi_\tau}$.
(See Fig.1).

\midinsert
\centerline{\epsffile{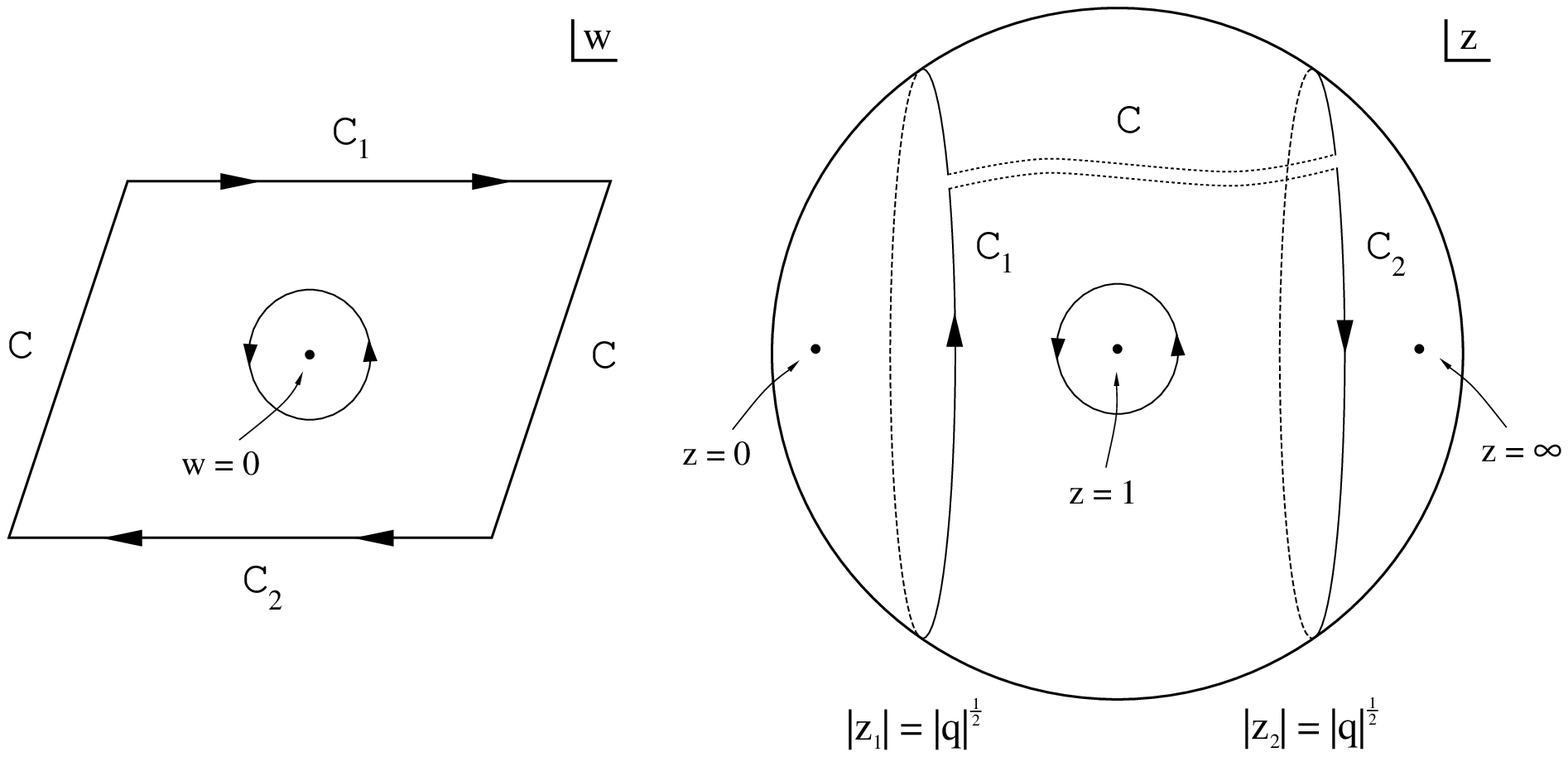}}
\centerline{Figure 1. The torus of modulus $\tau$ built by sewing
two punctures in the sphere $R_{123}^{\xi_\tau}$.}
\endinsert

The periodicity
$v_o(w) = v_o (w+1)$ in \yesp\ implies that the vector field
$v_o$ is well defined
on the annulus bounded by the curves $\C_1$ and $\C_2$,
corresponding to $|z_1| = |q|^{1/2}$ and $|z_2| = |q|^{1/2}$, respectively.
Since $v_o$ is analytic on the annulus minus the puncture,
it follows by contour deformation that the integral
around $w=0$ is equal to the sum of two integrals, one over $\C_1$ and the
other over $\C_2$:
$$\bra{R_{123}^{\xi_\tau}}\,\Bigl[\, -\oint_{\C_1} {dw\over 2\pi i}
b(w)v_o(w) \,
-  \oint_{\C_2} {dw\over 2\pi i} b(w) v_o(w)  \, \Bigr] \,
 q^{L_0^{(1)}} \bar q^{\overline L_0^{(1)}} \ket{R_{12}}\,. \eqn\oneev$$
Consider the antighost field $b(w)$ in the second integral and express
it in terms of $z_2$ coordinates,
$$b(w)\cdots
 q^{L_0^{(1)}} \bar q^{\overline L_0^{(1)}} \ket{R_{12}}
 =  \Bigl( {dz_2\over dw} \Bigr)^2 b(z_2) \cdots
 q^{L_0^{(1)}} \bar q^{\overline L_0^{(1)}} \ket{R_{12}}\,.\eqn\twoev$$
The antighost, living in the second state space,
can be taken all the way to the reflector
$\ket{R_{12}}$ at which point it can be re-expressed in
terms of the coordinate $z_1$,
$$b(w)\cdots
 q^{L_0^{(1)}} \bar q^{\overline L_0^{(1)}} \ket{R_{12}}
 = \Bigl( {dz_1\over dw} \Bigr)^2 \cdots q^{L_0^{(1)}}
\bar q^{\overline L_0^{(1)}} b (z_1) \ket{R_{12}}\,.\eqn\threeev$$
In order to bring the antighost
back to the left we use $q^{L_0} b(z)=q^2 b(qz)$,
$$b(w)\cdots
 q^{L_0^{(1)}} \bar q^{\overline L_0^{(1)}} \ket{R_{12}}
 =  \Bigl( {dz_1\over dw} \Bigr)^2  \, q^2 b (qz_1)\,\cdots
 q^{L_0^{(1)}} \bar q^{\overline L_0^{(1)}} \ket{R_{12}}\,,\eqn\fourev$$
and expressing the antighost back in $w$ coordinates we obtain,
$$ \Bigl( {dz_1\over dw} \Bigr)^2   \, q^2
\Bigl( {dw\over dz_1} \Bigr)^2_{qz_1} b (w+ \tau )\,
\cdots
 q^{L_0^{(1)}} \bar q^{\overline L_0^{(1)}} \ket{R_{12}}\,.\eqn\fiveev$$
The factor in front
of the antighost field is precisely unity. Returning to \oneev, we have,
$$\bra{R_{123}^{\xi_\tau}}\,\Bigl[\, -\oint_{\C_1} {dw\over 2\pi i}
 b(w) v_o(w) \,
-  \oint_{\C_2} {dw\over 2\pi i} b(w+\tau ) v_o(w)  \, \Bigr] \,
 q^{L_0^{(1)}} \bar q^{\overline L_0^{(1)}} \ket{R_{12}}\,. \eqn\sixev$$
Using $v_o(w+\tau)=1+v_o(w)$,
and a change of variable to move the contour from $\C_2$ to $-\C_1$,
$$-  \oint_{\C_2} {dw\over 2\pi i}\,b(w+\tau ) v_o (w) =
-  \oint_{\C_2} {dw\over 2\pi i}\, b(w+\tau )\, [ v_o(w + \tau) -1\, ]
=   \oint_{\C_1} {dw\over 2\pi i}\, b(w )\,[ v_o(w) -1\,]\,.\eqn\sevenev$$
The terms containing $v_o(w)$ in \sixev\ then cancel, leaving only,
$$\bra{R_{123}^{\xi_\tau}}\,\Bigl[\, - \oint_{\C_1} {dw\over 2\pi i}
b(w) \Bigr] \,
 q^{L_0^{(1)}} \bar q^{\overline L_0^{(1)}} \ket{R_{12}}\,. \eqn\sixev$$
This integral is  evaluated in the $z_1$ coordinate and gives simply
$(2\pi i) b_0^{(1)}$.
We therefore have
$$\bra{R_{123}^{\xi_\tau}} \,
q^{L_0^{(1)}} \bar q^{\overline L_0^{(1)}} \ket{R_{12}}\,
\oint_{\C_0} {dw\over 2\pi i} b(w) v_o(w) =
\bra{R_{123}^{\xi_\tau}} \,(-2\pi i\,b_0^{(1)})\,
q^{L_0^{(1)}} \bar q^{\overline L_0^{(1)}} \ket{R_{12}}\,. \eqn\sevtrtr$$
This was the desired result.

\refout
\end